%Paper: hep-ph/9209265
%From: "Fernando T.C. Brandt" <FBRANDT%USPIF%BRFAPESP.bitnet@uicvm.uic.edu>
%Date: Tue, 22 Sep 1992 18:00 BSC (-0300 C)

\tolerance 10000
\magnification\magstep1
\hsize=16.401truecm
\baselineskip=20pt
\newcount\eqnno \eqnno=0
\def\eqa{\global\advance\eqnno by 1 \eqno(1.\the\eqnno)}
\def\eqb{\global\advance\eqnno by 1 \eqno(2.\the\eqnno)}
\def\eqc{\global\advance\eqnno by 1 \eqno(3.\the\eqnno)}
\def\eapa{\global\advance\eqnno by 1 \eqno(A.\the\eqnno)}
\def\eapb{\global\advance\eqnno by 1 \eqno(B.\the\eqnno)}
\newcount\reff \reff=0
\def\ref#1{$[{#1}]$}
\def\pa{\vskip 6pt}
\def\PA{\vskip 18pt}

\hfill\break
\PA
\centerline{{\bf THE 3-GRAVITON VERTEX FUNCTION}}
\centerline{{\bf IN THERMAL QUANTUM GRAVITY}}
\pa
\pa
\centerline{F. T. Brandt and J. Frenkel}
\centerline{{\it Instituto de F\'\i sica, Universidade de S\~ao Paulo,}}
\centerline{{\it S\~ao Paulo, 01498 SP, Brasil}}
\PA
\centerline{September 1992}
\pa
\pa
\PA
\beginsection \centerline { }

\pa
The high temperature limit of the 3-graviton vertex function is studied in
thermal quantum gravity, to one loop order. The leading ($T^4$) contributions
arising from internal gravitons are calculated and shown to be twice the ones
associated with internal scalar particles, in correspondence with
the two helicity states of the graviton. The gauge invariance of
this result follows in consequence of the Ward and
Weyl identities obeyed by the thermal loops, which are verified explicitly.

\vfill\eject

\headline={\hfill\break}
\beginsection \centerline{I. INTRODUCTION}

\pa
There has been much work on quantum gravity at finite temperatures, which
are high compared with typical frequencies of the gravitational field. If the
temperature is well below the Planck scale, perturbation theory can be used
to calculate the n-graviton functions, with internal lines which correspond
to matter in thermal equilibrium. The functions for $n=1$ and $n=2$ have been
studied previously \ref{1,2} and show a leading $T^4$ behavior. Subsequently,
the work has been extended to the $n=3$ case, with a single loop of internal
scalar particles. Furthermore, it has been shown on general grounds, based
on the Ward and Weyl identities, that the partition function in a
gravitational field is determined uniquely in terms of the $O(\kappa)$
contributions $[3]$. Consequently, the contributions from internal scalars,
gluons and gravitons should be the same, up to simple numerical factors
which just count degrees of freedom.

\pa
The main purpose of this paper is to calculate the high-temperature limit
of the 3-graviton vertex function with a single loop of internal gravitons.
Besides verifying the general arguments presented in reference \ref{3}, this
study of thermal quantum gravity might offer new insights into the
general structure of the metric dependence of the partition function
at high temperature. The
calculation of the 3-graviton function is considerably more complicated than
that of the 3-gluon one \ref{4}. The method we use is an extension of that
in reference \ref{5}, in which the thermal Yang-Mills
n-point functions were related to
the forward-scattering amplitudes for the thermal Yang-Mills particles. This
method simplifies considerably the calculations in the present case, were we
consider the forward scattering amplitudes for the thermal gravitons in a
gravitational field.

\pa
In order to illustrate the method and to derive several results which will be
important later on, we first consider in Sec. II the graviton self-energy
function. We evaluate the leading temperature corrections and check the Ward
and Weyl identities which relate the self-energy function to the
energy-momentum tensor. In Sec. III, we derive the $T^4$ terms in the
3-graviton function, for the self-interacting thermal gravitons. We verify
the Ward and Weyl identities connecting the 3- and 2-point graviton functions.
The leading contributions have the same form as the one associated with
internal scalar particles, differing only by a factor of 2 in correspondence
with the two helicities of a physical graviton. Some mathematical
details which
arise during these calculations are given in the Appendices.

\beginsection \centerline{II. THE GRAVITON SELF-ENERGY}

\newcount\eqnno \eqnno=0
\pa
We consider here the leading high-temperature corrections to the graviton
self-energy in a space-time which is asymptotically flat. Hence, we expand
the metric tensor $g^{\mu\nu}$ in terms of the deviation from the
Minkowski metric $\eta^{\mu\nu}$ as follows
$$\sqrt{-g}g^{\mu\nu}\equiv
\tilde g^{\mu\nu}=\eta^{\mu\nu}+\kappa\phi^{\mu\nu},\eqb$$
where $\kappa=\sqrt{32\pi G}$ and $\phi^{\mu\nu}$ denotes the graviton field.
This enables us to evaluate perturbatively the thermal Green functions by
expanding the Einstein's Lagrangian written in the form \ref{6}:
$${\cal L}_{grav}={1\over 2\kappa^2}(\tilde g^{\rho\sigma}
\tilde g_{\lambda\mu}
\tilde g_{\kappa\nu}-{1\over 2}\tilde g^{\rho\sigma}\tilde g_{\mu\kappa}
\tilde g_{\lambda\nu}-
2\delta^\sigma_\kappa\delta^\rho_\lambda\tilde g_{\mu\nu})
\tilde g^{\mu\kappa}_{,\rho}\tilde g^{\lambda\nu}_{,\sigma}.\eqb$$
It is convenient to fix the gauge by choosing:
$${\cal L}_{fix}=-{1\over \kappa^2}(\partial_\mu\tilde g^{\mu\nu})^2,\eqb$$
which yields a contribution of the gravitational ghosts given by:
$${\cal L}_{ghost}=\bar\xi_\nu [\eta_{\nu\lambda} \partial^2-
\kappa(\phi_{\mu\nu ,\lambda\mu}-\phi_{\mu\rho}\eta_{\nu\lambda}\partial_\rho
\partial_\mu-\phi_{\mu\rho ,\mu}\eta_{\nu\lambda}\partial_\rho+
\phi_{\mu\nu ,\mu}\partial_\lambda)]\xi_\lambda.\eqb$$
The relevant Feynman rules following from the above Lagrangian density are
summarized in Appendix A.

\pa
The Feynman diagrams which contribute to the graviton self-energy function are
shown in Fig. 1. These graphs represent \ref{5} the forward scattering
amplitude of a thermal graviton with on-shell momenta
$q_\alpha=(q,\vec q)$ as indicated in
Fig. 2. This amplitude must be multiplied by the corresponding
Bose distribution function of the thermal graviton
and integrated over its 3-momentum $\vec q$. In this
way, we can express the thermal self-energy graviton function as:
$$\Gamma^2_{(\alpha\beta)(\mu\nu)}(k)=
{1\over 8 \pi^2}\int^\infty_0{q dq\over {\it e}^{q/T}-1}
\int {d\Omega\over 4\pi}\Gamma^2_{(\alpha\beta)(\mu\nu)}(k,q).\eqb$$

\pa
The high-temperature limit of the forward scattering amplitude
$\Gamma^2_{(\alpha\beta)(\mu\nu)}(k,q)$ is governed by
those parts of the Feynman
integrals which are superficially most divergent. To obtain these,
one needs to expand the Feynman denominator:
$${1\over k^2+2q\cdot k}={1\over 2 q\cdot k}-{k^2\over (2q\cdot k)^2}+\cdots
.\eqb$$
In this case there is a ``super-leading'' term of the form:
$${q_\alpha q_\beta q_\mu q_\nu\over q\cdot k},\eqb$$
which cancels between
the graphs of Fig. 2 and the corresponding crossed diagrams.
We are then left with the leading contributions which are functions of degree
two in $q$. Rescaling the null vector $q_\alpha$ by $q_\alpha=q Q_\alpha$,
where $Q_\alpha=(1,\hat Q)$ with $Q^2=0$, these can be expressed
in terms of the graviton energy density
$$\rho_g={1\over\pi^2}\int^\infty_0{q^3 dq\over {\it e}^{q/T}-1}=
{\pi^2 T^4\over 15}.\eqb$$
{}From the Feynman rules developed in Appendix A, we find that the
contributions to the Forward scattering amplitude
$\hat\Gamma^2_{(\alpha\beta)(\mu\nu)}(k,Q)$ associated
with the graphs in Fig. 2. are given respectively by:
$$\eqalign{
{1\over \kappa^2}\hat\Gamma^{2{\rm a}}_{(\alpha\beta)(\mu\nu)}(k,Q)&=
{{10\,{k}_{\nu }\,Q_{\alpha }\,Q_{\beta }\,Q_{\mu }}\over{{k} \cdot Q}}+
{{10\,{k}_{\mu }\,Q_{\alpha }\,Q_{\beta }\,Q_{\nu }}\over{{k} \cdot Q}}+
{{10\,{k}_{\beta }\,Q_{\alpha }\,Q_{\mu }\,Q_{\nu }}\over{{k} \cdot Q}}\cr
&+{{10\,{k}_{\alpha }\,Q_{\beta }\,Q_{\mu }\,Q_{\nu }}\over {{k}\cdot Q}}-
{{10\,{k}^{2}\,Q_{\alpha}\,Q_{\beta}\,Q_{\mu }\,Q_{\nu }}\over
{{{(k\cdot Q)}^2}}}\cr
&-10\,Q_{\mu }\,Q_{\nu}\,\eta_{\alpha\beta }-
10\,Q_{\alpha }\,Q_{\beta}\,\eta_{\mu\nu},}\eqb$$
$$\eqalign{
{1\over \kappa^2}\hat\Gamma^{2{\rm b}}_{(\alpha\beta)(\mu\nu)}(k,Q)=&
-{{8\,{k}_{\nu }\,Q_{\alpha }\,Q_{\beta }\,Q_{\mu }}\over{{k} \cdot Q}}-
 {{8\,{k}_{\mu }\,Q_{\alpha }\,Q_{\beta }\,Q_{\nu }}\over{{k} \cdot Q}}-
 {{8\,{k}_{\beta }\,Q_{\alpha }\,Q_{\mu }\,Q_{\nu }}\over{{k} \cdot Q}}\cr
&-{{8\,{k}_{\alpha }\,Q_{\beta }\,Q_{\mu }\,Q_{\nu }}\over {{k} \cdot Q}}+
 {{8\,{k}^{2}\,Q_{\alpha }\,Q_{\beta }\,Q_{\mu }\,Q_{\nu }}
\over{{(k \cdot Q)^2}}},}\eqb$$
$${1\over \kappa^2}\hat\Gamma^{2{\rm c}}_{(\alpha\beta)(\mu\nu)}(k,Q)=
10(\eta_{\alpha\beta}Q_\mu Q_\nu+\eta_{\mu\nu}Q_\alpha Q_\beta).\eqb$$
Note that equations (2.9) and (2.11) contain terms
involving the Minkowski metric tensor.
However, such terms cancel out in the total sum which gives twice the
contribution one gets from a single loop of internal scalar particles:
$$\eqalign{{1\over \kappa^2}
\hat\Gamma^{2}_{(\alpha\beta)(\mu\nu)}(k,Q)=&2\left(
{{{k}_{\nu }\,Q_{\alpha }\,Q_{\beta }\,Q_{\mu }}\over{{k} \cdot Q}}+
{{{k}_{\mu }\,Q_{\alpha }\,Q_{\beta }\,Q_{\nu }}\over{{k} \cdot Q}}+
{{{k}_{\beta }\,Q_{\alpha }\,Q_{\mu }\,Q_{\nu }}\over{{k} \cdot Q}}\right.\cr
&+\left.{{{k}_{\alpha }\,Q_{\beta }\,Q_{\mu }\,Q_{\nu }}\over {{k}\cdot Q}}-
{{{k}^{2}\,Q_{\alpha}\,Q_{\beta}\,Q_{\mu }\,Q_{\nu }}\over{{{(k\cdot Q)}^2}}}
\right).}\eqb$$
This result is expected from the Ward identity relating
the self-energy function
to the energy-momentum tensor:
$$(2\eta^{\alpha\lambda}{k_1}^\beta-\eta^{\alpha\beta}{k_1}^\lambda)
\Gamma^2_{(\alpha\beta)(\mu\nu)}(k_1)=
\eta^{\beta\lambda}(\delta^\alpha_\mu {k_1}_\nu+\delta^\alpha_\nu {k_1}_\mu)
{T_{\alpha\beta}\over 2},\eqb$$
and from the invariance under Weyl transformations which requires:
$$\eta^{\alpha\beta}\Gamma^2_{(\alpha\beta)(\mu\nu)}(k)=
{T_{\mu\nu}\over 2}.\eqb$$
As shown in reference \ref{3}, these relations fix uniquely the self-energy
function in terms of the energy-momentum tensor. Since the contributions
to $T_{\mu\nu}$ from internal scalars and gravitons are all the same \ref{2}
apart from simple numerical factors counting the degrees of freedom, the result
expressed by Eq. (2.12) should be expected.

\beginsection \centerline{III. THE 3-GRAVITON VERTEX FUNCTION}

\newcount\eqnno \eqnno=0
\pa
We now turn to the leading temperature corrections of the 3-point graviton
function. The thermal loop diagrams which are relevant to our discussion are
shown in Fig. 3. and the corresponding forward scattering amplitudes are
represented in Fig. 4. In the high temperature limit we require
large values of
momenta $q_\alpha=(q,\vec q)$ associated with the thermal graviton.
Since the
thermal particle is on shell, each Feynman denominator in the diagrams
in Fig. 4 has the form:
$$(2 q\cdot k+k^2)^{-1}~~~{\rm with}~~~k=\sum k_i,\eqc$$
the sum being over some set of indices i. We may expand each denominator in
powers of $(k^2/2q\cdot k)$, and also the numerators in powers of
$k_{i\mu}/q$. The first term has a denominator which is quadratic in
$(q\cdot k)^{-1}$, and a numerator with the single tensor structure
$q_\alpha q_\beta q_\mu q_\nu q_\rho q_\sigma$.
However, these terms cancel when all graphs are added, as a
consequence of the eikonal identity:
$$(q\cdot k_1)^{-1}(q\cdot k_2)^{-1}+(q\cdot k_1)^{-1}(q\cdot k_3)^{-1}+
(q\cdot k_2)^{-1}(q\cdot k_3)^{-1}=0,\eqc$$
since $k_1+k_2+k_3=0$.
The next terms are down by a power of $k_i/q$, being individually
``super-leading''. Also these turn out to cancel out by a combination of
the eikonal identity and the requirement of Bose symmetry.

\pa
With the super-leading terms out of the way, we now consider the leading
contributions. To this end we take the integrand to one further term in powers
of the external momenta. Then the leading terms will become homogeneous
functions of $q$ of degree 2. We proceed as in the previous section, rescaling
$q_\alpha$ by a factor of $q$: $q_\alpha=q Q_\alpha$, and integrating over $q$.
Then the leading behavior of the 3-graviton vertex can be expressed as:
$$\Gamma^3_{(\alpha\beta)(\mu\nu)(\rho\sigma)}(k_1,k_2,k_3)=
{\rho_g\over 8}\int {d\Omega\over 4\pi}
\hat\Gamma^3_{(\alpha\beta)(\mu\nu)(\rho\sigma)}(k_1,k_2,k_3,Q),\eqc$$
where the graviton energy density $\rho_g$ (equation (2.8)) is proportional
to $T^4$. Using the Feynman rules given in Appendix A, we perform the
necessary algebra in order to find the leading contributions to the forward
scattering
amplitude  $\hat\Gamma^3_{(\alpha\beta)(\mu\nu)(\rho\sigma)}(k_1,k_2,k_3,Q)$.
The calculation is considerably involved, requiring the use of vast algebraic
manipulations. The corresponding corrections, associated with the graphs
shown in Fig. 4., can be written in the form given in Appendix B. Here we
discuss only the main features of these complex algebraic structures.

\pa
The contributions from diagram (4a), involving only three-graviton vertices
has the general from (see Eq. B1):
$$\eqalign{
\hat\Gamma^{3{\rm a}}_{(\alpha\beta)(\mu\nu)(\rho\sigma)}(k_1,k_2,k_3,Q)&=
A Q_\alpha Q_\beta Q_\mu Q_\nu Q_\rho Q_\sigma+
B^1_\alpha Q_\beta Q_\mu Q_\nu Q_\rho Q_\sigma
+(\alpha \leftrightarrow \beta)\cr
&+B^2_\mu Q_\alpha Q_\beta Q_\nu Q_\rho Q_\sigma + \cdots\cr
&+C^1_{\alpha\beta}Q_\mu Q_\nu Q_\rho Q_\sigma + \cdots +
D^{1 2}_{\alpha\mu}Q_\beta Q_\nu Q_\rho Q_\sigma + \cdots \cr
&+E^{1 2}_\alpha\eta_{\mu\nu}Q_\beta Q_\rho Q_\sigma + \cdots
+F^{1 2}_\alpha\eta_{\beta\mu}Q_\nu Q_\rho Q_\sigma + \cdots \cr
&+G^1_\alpha\eta_{\mu\sigma}Q_\beta Q_\nu Q_\rho + \cdots
+ H^1\eta_{\alpha\beta}Q_\mu Q_\nu Q_\rho Q_\sigma + \cdots \cr
&+I^{1 2}\eta_{\alpha\mu}Q_\beta Q_\nu Q_\rho Q_\sigma + \cdots\cr
&+J\left(\eta_{\alpha\beta}\eta_{\mu\nu}Q_\rho Q_\sigma+\cdots\right)
+J'\left(\eta_{\alpha\beta}\eta_{\mu\rho}Q_\nu Q_\sigma+\cdots\right)\cr
&+J''\left(\eta_{\alpha\mu}\eta_{\beta\nu}Q_\rho Q_\sigma+\cdots\right)+
J'''\left(\eta_{\alpha\mu}\eta_{\beta\rho}Q_\nu Q_{\sigma}+\cdots\right).}
\eqc$$
Here $J$, $J'$, $J''$, $J'''$ are constants, $A$, $H^i$, $I^{i j}$ are
scalars functions of $Q$ and the external momenta, $B^i_\alpha$,
$E^{i j}_\alpha$, $F^{i j}_\alpha$, $G^i_\alpha$ are vector functions of Q
and the external momenta not proportional to $Q_\alpha$, and
$C^i_{\alpha\beta}$, $D^{i j}_{\alpha\beta}$ are tensor functions containing
neither $Q_\alpha$ nor $Q_\beta$. The ellipses denote the addition of as many
terms as are necessary to symmetrize under $(\alpha\leftrightarrow\beta)$,
$(\mu\leftrightarrow\nu)$, $(\rho\leftrightarrow\sigma)$ and under
the permutations of $(k_1,~\alpha,~\beta)$, $(k_2,~\mu,~\nu)$,
$(k_3,~\rho,~\sigma)$. Note that each term is a function of degree two in $Q$
and of zero degree in the external momenta.

\pa
The contribution from the ghost particles in Fig. (4b) is less complicated
algebraically than the previous one, since the terms proportional to the
Minkowski metric $\eta$ are absent (see Eq. B2). On the other hand, the
corrections arising from graph (4c) give, as shown in Eq. (B3), only terms
which contain explicitly the Minkowski metric $\eta$. Finally, the
contributions
associated with diagram (4d), which involves the 5-graviton coupling, have a
structure proportional to $\eta\otimes\eta$, as can be seen from equation (B4).

\pa
However, the terms depending explicitly on the Minkowski metric cancel
when all graphs are added. The final result for the forward scattering
amplitude is rather simple and can be written in the form (cf. Eq. (B5)):
$$\eqalign{\hat\Gamma^3_{(\alpha\beta)(\mu\nu)(\rho\sigma)}(k_1,k_2,k_3,Q)&=
\hat A Q_\alpha Q_\beta Q_\mu Q_\nu Q_\rho Q_\sigma
+\hat B^1_\alpha Q_\beta Q_\mu Q_\nu Q_\rho Q_\sigma
+(\alpha\leftrightarrow\beta)\cr
&+\hat B^2_\mu Q_\alpha Q_\beta Q_\nu Q_\rho Q_\sigma+\cdots
+\hat C^1_{\alpha\beta}Q_\mu Q_\nu Q_\rho Q_\sigma+\cdots\cr
&+\hat D^{1 2}_{\alpha\mu}Q_\beta Q_\nu Q_\rho Q_\sigma+\cdots,}\eqc$$
where $Q_\alpha=(1,\hat Q)$, and
$$\hat A={k_1^2 k_2\cdot k_3\over (k_1\cdot Q)^2 k_2\cdot Q k_3\cdot Q}+
(cyclic~permutations),\eqc$$
$$\hat B^1_\alpha=
-{k_2\cdot k_3 k_{1\alpha}\over k_1\cdot Q k_2\cdot Q k_3\cdot Q}
-{k_3^2 k_{2\alpha}\over  k_2\cdot Q (k_3\cdot Q)^2}
-{k_2^2 k_{3\alpha}\over (k_2\cdot Q)^2 k_3\cdot Q},\eqc$$
$$\hat C^1_{\alpha\beta}={k_{2\alpha}k_{3\beta}+k_{2\beta}k_{3\alpha}\over
k_2\cdot Q k_3\cdot Q},\eqc$$
$$\hat D^{1 2}_{\alpha\mu}={k_{1\alpha}k_{3\mu}\over
k_1\cdot Q k_3\cdot Q}+
{k_{2\mu}k_{3\alpha}\over
k_2\cdot Q k_3\cdot Q}.\eqc$$
The other terms in (3.5) containing the coefficients $\hat B^i$, $\hat C^i$
and $\hat D^{i j}$ can be obtained respectively from (3.7), (3.8)
and (3.9) by symmetry.

\pa
The important point about (3.5) is that, under Lorentz transformations in the
asymptotic Minkowski space, it is a covariant function of the null vector
$Q_\alpha$. We remark that
equation (3.5) is a homogeneous expression of $Q$ of degree 2
and of zero degree in the external momenta. It gives precisely twice the
contribution arising from a single loop of internal scalar particles.

\pa
In order to understand this result, we now consider the Ward identity which
follows as a consequence of the invariance under general coordinate
transformations. This relation, which connects the 3-graviton vertex
(Eqs. (3.3) and (3.5)) to the
self energy function (Eqs. (2.5), (2.8) and (2.12)),
is verified explicitly by:
$$\eqalign{(2\eta^{\alpha\lambda}k_1^\beta-\eta^{\alpha\beta}k_1^\lambda)
\Gamma^3_{(\alpha\beta)(\mu\nu)(\rho\sigma)}(k_1,k_2,k_3)=&
[\eta^{\beta\lambda}(\delta^\alpha_\mu {k_1}_\nu+\delta^\alpha_\nu {k_1}_\mu)
+\delta^\alpha_\mu\delta^\beta_\nu {k_3}^\lambda]
\Gamma^2_{(\alpha\beta)(\rho\sigma)}(k_3)\cr
&+(k_2,~\mu,~\nu)\leftrightarrow(k_3,~\rho,~\sigma).}\eqc$$
Since the thermal corrections from the ghost-ghost-graviton vertex
functions are sub-leading \ref{3}, these terms do not contribute to
Eq. (3.10). Consequently, the Ward identity has the same form as in a physical
gauge, indicating that the leading $T^4$ contributions are gauge
independent. This property can be ascertained by considering also the Weyl
invariance of the theory, which reflects its invariance under scale
transformations. From this, we verify the trace identity:
$$\eta^{\alpha\beta}\Gamma^3_{(\alpha\beta)(\mu\nu)(\rho\sigma)}(k_1,k_2,k_3)=
 \Gamma^2_{(\mu\nu)(\rho\sigma)}(k_2)
+\Gamma^2_{(\rho\sigma)(\mu\nu)}(k_3).\eqc$$
It has been argued in \ref{3} that the use of the
Ward and Weyl identities,
together with the structure indicated by equation (3.5), are sufficient to fix
uniquely the 3-point vertex in terms of the self-energy function. As we
have seen, the same identities determine the 2-point function in terms of the
energy-momentum tensor $T^{\mu\nu}$. Hence, these relations are sufficient to
determine uniquely the 3-point graviton vertex in terms of $T^{\mu\nu}$.

\pa
It is well known \ref{7} that in gauge theories, the thermal corrections to
the energy-momentum tensor should be gauge invariant
quantities. The contributions to $T^{\mu\nu}$ from internal scalars
and gravitons are the same,
up to a factor of 2 which counts the graviton degrees of freedom. From the
above arguments, it then follows that the leading $T^4$ contributions to the
3-graviton vertex function, from a single loop of internal gravitons, must be
gauge invariant. Furthermore, these corrections should be twice
the ones arising from an internal loop of scalar particles,
as expected from the two helicity states of the graviton.

\beginsection \centerline {ACKNOWLEDGMENTS}

\pa
We would like to thank ${\rm CNP}_{\rm q}$ and FAPESP (Brasil) for support.
The authors are indebted to Prof. J.C. Taylor for a helpful
correspondence.

\beginsection \centerline {APPENDIX A}

\newcount\eqnno \eqnno=0
\pa
In this appendix we derive the Feynman rules which are relevant for the
perturbative study of quantum gravity. To this end, we write the total
Lagrangian density obtained from the sum of (2.2), (2.3) and (2.4) as:
$${\cal L}=\sum^\infty_{j=2}\kappa^{j-2}{\cal L}_{(j)}
+ {\cal L}_{ghost},\eapa$$
where the terms in the series are given by increasing powers of the
graviton field defined in Eq. (2.1). Noticing that:
$$\tilde g_{\mu\nu}=\eta_{\mu\nu}-\kappa\phi_{\mu\nu}
                   +\kappa^2\phi_{\mu\alpha}\phi_{\alpha\nu}
                   -\kappa^3\phi_{\mu\alpha}\phi_{\alpha\beta}\phi_{\beta\nu}
                   +O(\kappa^4),\eapa$$
and inverting the tensor which multiplies $\phi_{\alpha\beta}\phi_{\mu\nu}$,
we find that the graviton propagator in momentum space is given by
$$D^{grav.}_{(\alpha\beta)(\mu\nu)}(k)={1\over
 2 k^2}(\eta_{\alpha\mu}\eta_{\beta\nu}+\eta_{\alpha\nu}\eta_{\beta\mu}
                                       -\eta_{\alpha\beta}\eta_{\mu\nu}).
\eapa$$
In the same way,
the quadratic part in the ghost field leads to the ghost propagator
$$D^{ghost}_{\alpha\beta}(k)={\eta_{\alpha\beta}\over k^2}.\eapa$$
The ghost-ghost-graviton vertex can be easily derived from Eq. (2.4). We
obtain in momentum space that:
$$\Gamma^{ghost}_{(\alpha\beta)(\mu)(\nu)}(k_1,k_2,k_3)={\kappa\over 2}
[\eta_{\mu\nu}({k_2}_\alpha {k_3}_\beta+{k_3}_\alpha {k_2}_\beta)-
       (\eta_{\alpha\mu}{k_1}_\beta +
        \eta_{\beta\mu}{k_1}_\alpha ){k_2}_\nu].\eapa$$
The graviton-graviton couplings in momentum space represent an algebraic
problem for which we have used the computer program Mathematica. We find for
these couplings the following expressions:
\vfill\eject
\centerline{\bf The three graviton coupling}

$$\eqalign{{4\over\kappa}\times &
\Gamma^3_{(\alpha\beta)(\mu\nu)(\rho\sigma)}(k_1,k_2,k_3)=\cr
\left[\right.&\left.\right.
-4\,{k_2}_{\rho }\,{k_3}_{\nu }\,\eta_{\alpha\mu }\,\eta_{\beta\sigma }
-{k_2}\cdot{k_3}\,\eta_{\alpha\rho }\,\eta_{\beta\sigma}\,\eta_{\mu\nu}\cr
&+2\,{k_2}\cdot {k_3}\,\eta_{\alpha\nu}\,
\eta_{\beta\sigma}\,\eta_{\mu\rho} +2\,
{k_2} \cdot {k_3}\,\eta_{\alpha  \mu }\,\eta_{\beta  \rho }\,
\eta_{\nu  \sigma }\cr
&-2\,{k_2}_{\alpha}\,{k_3}_{\beta}\,\eta_{\mu\rho}\,\eta_{\nu\sigma }
- {k_2}\cdot{k_3}\,\eta_{\alpha\mu }\,\eta_{\beta\nu}\,
\eta_{\rho\sigma }\cr
&+{k_2}_{\alpha}\,{k_3}_{\beta}\,\eta_{\mu\nu}\,\eta_{\rho\sigma} \cr
&+(symmetrization~under~(\alpha\leftrightarrow\beta),~
                        (\mu\leftrightarrow\nu),~
                        (\rho\leftrightarrow\sigma))~\left.\right]\cr
&+\left(permutations~of~(k_1,\alpha,\beta),~(k_2,\mu,\nu),~(k_3,\rho,\sigma)~
\right).\cr}\eapa$$
\PA
\centerline{\bf The four graviton coupling}

$$\eqalign{{4\over\kappa^2}\times &
\Gamma^4_{(\alpha\beta)(\mu\nu)(\rho\sigma)(\lambda\tau)}(k_1,k_2,k_3,k_4)=\cr
\left[\right.&\left.\right.
-2\,{k_3}_{\mu }\,{k_4}_{\nu }\,\eta_{\alpha  \sigma }\,
\eta_{\beta  \tau }\,\eta_{\lambda  \rho }
+ {k_3}_{\mu }\,{k_4}_{\nu }\,\eta_{\alpha  \rho }\,
   \eta_{\beta  \sigma }\,\eta_{\lambda  \tau } \cr
&-{k_3} \cdot {k_4}\,\eta_{\alpha  \rho }\,\eta_{\beta  \nu }\,
   \eta_{\lambda  \tau }\,\eta_{\mu  \sigma }
-4\,{k_3}_{\lambda }\,{k_4}_{\sigma }\,\eta_{\alpha  \rho }\,
   \eta_{\beta  \nu }\,\eta_{\mu  \tau } \cr
&+2\,{k_3} \cdot {k_4}\,\eta_{\alpha  \sigma }\,\eta_{\beta  \nu }\,
   \eta_{\lambda  \rho }\,\eta_{\mu  \tau }
- {k_3} \cdot {k_4}\,\eta_{\alpha  \rho }\,\eta_{\beta  \sigma }\,
   \eta_{\lambda  \mu }\,\eta_{\nu  \tau } \cr
&+ 2\,{k_3} \cdot {k_4}\,\eta_{\alpha  \rho }\,\eta_{\beta  \lambda }\,
   \eta_{\mu  \sigma }\,\eta_{\nu  \tau }
+   {k_3}_{\mu }\,{k_4}_{\nu }\,\eta_{\alpha  \lambda }\,\eta_{\beta  \tau }\,
   \eta_{\rho  \sigma } \cr
&- {k_3} \cdot {k_4}\,\eta_{\alpha  \lambda }\,
   \eta_{\beta  \nu }\,\eta_{\mu  \tau }\,\eta_{\rho  \sigma }
- 2\,{k_3}_{\mu }\,{k_4}_{\nu }\,\eta_{\alpha  \rho }\,
   \eta_{\beta  \lambda }\,\eta_{\sigma  \tau } \cr
&+ 2\,{k_3} \cdot {k_4}\,\eta_{\alpha  \rho }\,\eta_{\beta  \nu }\,
   \eta_{\lambda  \mu }\,\eta_{\sigma  \tau }\cr
&+(symmetrization~under~(\alpha\leftrightarrow\beta),~
                               (\mu\leftrightarrow\nu),~
                               (\rho\leftrightarrow\sigma),~
                               (\lambda\leftrightarrow\tau))~\left.\right]\cr
&+\left(permutations~of~(k_1,\alpha,\beta),~(k_2,\mu,\nu),~(k_3,\rho,\sigma),~
(k_4,\lambda,\tau)~\right).\cr}\eapa$$

\vfill\eject
\centerline{\bf The five graviton coupling}

$$\eqalign{{4\over\kappa^3}\times &
\Gamma^5_{(\alpha\beta)(\mu\nu)(\rho\sigma)(\lambda\tau)
(\gamma\delta)}(k_1,k_2,k_3,k_4,k_5)=\cr
\left[\right.&\left.\right.
-2\,{k_4}_{\rho }\,{k_5}_{\sigma }
\,\eta_{\alpha \mu }\,\eta_{\beta\tau }\,
   \eta_{\delta  \nu }\,\eta_{\gamma  \lambda } +
  {k_4}_{\rho }\,{k_5}_{\sigma }\,\eta_{\alpha  \tau }\,
   \eta_{\beta  \lambda }\,\eta_{\delta  \nu }\,\eta_{\gamma  \mu }\cr
&+ {k_4}_{\rho }\,{k_5}_{\sigma }\,\eta_{\alpha  \mu }\,\eta_{\beta  \tau }\,
   \eta_{\delta  \gamma }\,\eta_{\lambda  \nu }
-4\,{k_4}_{\gamma }\,{k_5}_{\tau }\,\eta_{\alpha  \mu }\,
   \eta_{\beta  \rho }\,\eta_{\delta  \sigma }\,\eta_{\lambda  \nu }\cr
&-  2\,{k_4}_{\rho }\,{k_5}_{\sigma }\,\eta_{\alpha  \mu }\,
   \eta_{\beta  \gamma }\,\eta_{\delta  \tau }\,\eta_{\lambda  \nu }
- 2\,{k_4}_{\rho }\,{k_5}_{\sigma }\,\eta_{\alpha  \tau }\,
   \eta_{\beta  \delta }\,\eta_{\gamma  \mu }\,\eta_{\lambda  \nu } \cr
&-{k_4} \cdot {k_5}\,\eta_{\alpha  \mu }\,\eta_{\beta  \tau }\,
   \eta_{\delta  \sigma }\,\eta_{\gamma  \rho }\,\eta_{\lambda  \nu }
+2\,{k_4} \cdot {k_5}\,\eta_{\alpha  \mu }\,\eta_{\beta  \rho }\,
   \eta_{\delta  \tau }\,\eta_{\gamma  \nu }\,\eta_{\lambda  \sigma } \cr
&+2\,{k_4} \cdot {k_5}\,\eta_{\alpha  \mu }\,\eta_{\beta  \tau }\,
   \eta_{\delta  \nu }\,\eta_{\gamma  \rho }\,\eta_{\lambda  \sigma }
+{k_4}_{\rho }\,{k_5}_{\sigma }\,\eta_{\alpha  \mu }\,\eta_{\beta  \gamma }\,
   \eta_{\delta  \nu }\,\eta_{\lambda  \tau } \cr
&-{k_4} \cdot {k_5}\,\eta_{\alpha  \mu }\,\eta_{\beta  \rho }\,
   \eta_{\delta  \sigma }\,\eta_{\gamma  \nu }\,\eta_{\lambda  \tau }
+  2\,{k_4} \cdot {k_5}\,\eta_{\alpha  \mu }\,\eta_{\beta  \rho }\,
   \eta_{\delta  \sigma }\,\eta_{\gamma  \lambda }\,\eta_{\nu  \tau } \cr
&+2\,{k_4} \cdot {k_5}\,\eta_{\alpha  \mu }\,\eta_{\beta  \gamma }\,
   \eta_{\delta  \sigma }\,\eta_{\lambda  \nu }\,\eta_{\rho  \tau }
- {k_4} \cdot {k_5}\,\eta_{\alpha  \mu }\,\eta_{\beta  \gamma }\,
   \eta_{\delta  \nu }\,\eta_{\lambda  \sigma }\,\eta_{\rho  \tau } \cr
&- {k_4} \cdot {k_5}\,\eta_{\alpha  \mu }\,\eta_{\beta  \rho }\,
   \eta_{\delta  \gamma }\,\eta_{\lambda  \nu }\,\eta_{\sigma  \tau }\cr
&+(symmetrization~under~(\alpha\leftrightarrow\beta),~
                               (\mu\leftrightarrow\nu),~
                               (\rho\leftrightarrow\sigma),~
                               (\lambda\leftrightarrow\tau),~
                               (\gamma\leftrightarrow\delta))~\left.\right]\cr
&+\left(permutations~of~(k_1,\alpha,\beta),~(k_2,\mu,\nu),~(k_3,\rho,\sigma),~
(k_4,\lambda,\tau),~(k_5,\gamma,\delta)~\right).\cr}\eapa$$
As usual, we have energy-momentum conservation at the vertices, where all
momenta are defined to be inwards.

\beginsection \centerline {APPENDIX B}

\newcount\eqnno \eqnno=0

\pa
Here we present the complete expressions for the contributions to the forward
scattering amplitude which are associated with the diagrams in Fig. 4.
{}From the Feynman rules developed in Appendix A we obtain,
after a vast algebraic manipulation, the following results:

\vfill\eject
$$\eqalign{&{
1\over \kappa^3}\hat\Gamma^{3a}_{(\alpha\beta)(\mu\nu)(\rho\sigma)}
(k_1,k_2,k_3,Q)= \cr &
\left[\right.
-{{7\,{k_1}_{\rho }\,{k_1}_{\sigma }\,Q_{\alpha }\,Q_{\beta }\,
      Q_{\mu }\,Q_{\nu }}\over {{k_1}\cdot Q\,{k_2}\cdot Q}} -
  {{4\,{k_1}_{\rho }\,{k_2}_{\sigma }\,Q_{\alpha }\,Q_{\beta }\,
      Q_{\mu }\,Q_{\nu }}\over {{k_1}\cdot Q\,{k_2}\cdot Q}} -
  {{7\,{k_2}_{\rho }\,{k_2}_{\sigma }\,Q_{\alpha }\,Q_{\beta }\,
      Q_{\mu }\,Q_{\nu }}\over {{k_1}\cdot Q\,{k_2}\cdot Q}} \cr & +
  {{12\,{k_1}_{\rho }\,{k_2}_{\mu }\,Q_{\alpha }\,Q_{\beta }\,Q_{\nu }\,
      Q_{\sigma }}\over {{k_1}\cdot Q\,{k_2}\cdot Q}} +
  {{8\,{k_1}_{\mu }\,{k_2}_{\rho }\,Q_{\alpha }\,Q_{\beta }\,Q_{\nu }\,
      Q_{\sigma }}\over {{k_1}\cdot Q\,{k_2}\cdot Q}} +
  {{8\,{k_1}_{\rho }\,{k_2}_{\alpha }\,Q_{\beta }\,Q_{\mu }\,Q_{\nu }\,
      Q_{\sigma }}\over {{k_1}\cdot Q\,{k_2}\cdot Q}} \cr & +
  {{12\,{k_1}_{\alpha }\,{k_2}_{\rho }\,Q_{\beta }\,Q_{\mu }\,Q_{\nu }\,
      Q_{\sigma }}\over {{k_1}\cdot Q\,{k_2}\cdot Q}} -
  {{5\,{k_1}_{\rho }\,k_2^2\,Q_{\alpha }\,Q_{\beta }\,Q_{\mu }\,
      Q_{\nu }\,Q_{\sigma }}\over {{k_1}\cdot Q\,{(k_2\cdot Q)^2}}} +
{{5\,k_1^2\,{k_1}_{\rho }\,Q_{\alpha }\,Q_{\beta }\,Q_{\mu }\,
      Q_{\nu }\,Q_{\sigma }}\over {{(k_1\cdot Q)^2}\,{k_2}\cdot Q}} \cr & +
 {{5\,k_2^2\,{k_2}_{\rho }\,Q_{\alpha }\,Q_{\beta }\,Q_{\mu }\,
      Q_{\nu }\,Q_{\sigma }}\over {{k_1}\cdot Q\,{(k_2\cdot Q)^2}}}
   - {{5\,k_1^2\,{k_2}_{\rho }\,Q_{\alpha }\,Q_{\beta }\,Q_{\mu }\,
      Q_{\nu }\,Q_{\sigma }}\over {{(k_1\cdot Q)^2}\,{k_2}\cdot Q}}
   - {{7\,{k_2}_{\mu }\,{k_2}_{\nu }\,Q_{\alpha }\,Q_{\beta }\,
      Q_{\rho }\,Q_{\sigma }}\over {{k_1}\cdot Q\,{k_2}\cdot Q}} \cr & -
  {{8\,{k_1}_{\mu }\,{k_2}_{\alpha }\,Q_{\beta }\,Q_{\nu }\,Q_{\rho }\,
      Q_{\sigma }}\over {{k_1}\cdot Q\,{k_2}\cdot Q}} +
  {{8\,{k_1}_{\alpha }\,{k_2}_{\mu }\,Q_{\beta }\,Q_{\nu }\,Q_{\rho }\,
      Q_{\sigma }}\over {{k_1}\cdot Q\,{k_2}\cdot Q}} +
  {{5\,k_2^2\,{k_2}_{\mu }\,Q_{\alpha }\,Q_{\beta }\,Q_{\nu }\,
      Q_{\rho }\,Q_{\sigma }}\over {{k_1}\cdot Q\,{(k_2\cdot Q)^2}}}
    \cr & -
 {{5\,k_1^2\,{k_2}_{\mu }\,Q_{\alpha }\,Q_{\beta }\,Q_{\nu }\,
      Q_{\rho }\,Q_{\sigma }}\over {{(k_1\cdot Q)^2}\,{k_2}\cdot Q}}
    - {{7\,{k_1}_{\alpha }\,{k_1}_{\beta }\,Q_{\mu }\,Q_{\nu }\,
      Q_{\rho }\,Q_{\sigma }}\over {{k_1}\cdot Q\,{k_2}\cdot Q}} -
  {{5\,{k_1}_{\alpha }\,k_2^2\,Q_{\beta }\,Q_{\mu }\,Q_{\nu }\,
      Q_{\rho }\,Q_{\sigma }}\over {{k_1}\cdot Q\,{(k_2\cdot Q)^2}}}
    \cr & +
 {{5\,k_1^2\,{k_1}_{\alpha }\,Q_{\beta }\,Q_{\mu }\,Q_{\nu }\,
      Q_{\rho }\,Q_{\sigma }}\over {{(k_1\cdot Q)^2}\,{k_2}\cdot Q}}
    - {{5\,{k_2^4}\,Q_{\alpha }\,Q_{\beta }\,Q_{\mu }\,Q_{\nu }\,
      Q_{\rho }\,Q_{\sigma }}\over
    {2\,{k_1}\cdot Q\,{(k_2\cdot Q)^3}}} +
  {{5\,k_1^2\,k_2^2\,Q_{\alpha }\,Q_{\beta }\,Q_{\mu }\,
      Q_{\nu }\,Q_{\rho }\,Q_{\sigma }}\over
    {2\,{(k_1\cdot Q)^2}\,{(k_2\cdot Q)^2}}} \cr & -
  {{5\,{k_1^4}\,Q_{\alpha }\,Q_{\beta }\,Q_{\mu }\,Q_{\nu }\,
      Q_{\rho }\,Q_{\sigma }}\over
    {2\,{(k_1\cdot Q)^3}\,{k_2}\cdot Q}} \cr & -
  {{5\,{k_1}_{\rho }\,Q_{\mu }\,Q_{\nu }\,Q_{\sigma }\,
      \eta_{\alpha\beta }}\over {{k_1}\cdot Q}} +
  {{2\,{k_1}_{\rho }\,Q_{\mu }\,Q_{\nu }\,Q_{\sigma }\,
      \eta_{\alpha\beta }}\over {{k_2}\cdot Q}} +
  {{2\,{k_2}_{\rho }\,Q_{\mu }\,Q_{\nu }\,Q_{\sigma }\,
      \eta_{\alpha\beta }}\over {{k_1}\cdot Q}} \cr & -
  {{5\,{k_2}_{\rho }\,Q_{\mu }\,Q_{\nu }\,Q_{\sigma }\,
      \eta_{\alpha\beta }}\over {{k_2}\cdot Q}} +
  {{3\,{k_1}_{\mu }\,Q_{\nu }\,Q_{\rho }\,Q_{\sigma }\,
      \eta_{\alpha\beta }}\over {{k_1}\cdot Q}} +
  {{2\,{k_2}_{\mu }\,Q_{\nu }\,Q_{\rho }\,Q_{\sigma }\,
      \eta_{\alpha\beta }}\over {{k_1}\cdot Q}} \cr & -
  {{5\,{k_2}_{\mu }\,Q_{\nu }\,Q_{\rho }\,Q_{\sigma }\,
      \eta_{\alpha\beta }}\over {{k_2}\cdot Q}} +
  {{k_1^2\,Q_{\mu }\,Q_{\nu }\,Q_{\rho }\,Q_{\sigma }\,
      \eta_{\alpha\beta }}\over {{(k_1\cdot Q)^2}}} +
  {{5\,k_2^2\,Q_{\mu }\,Q_{\nu }\,Q_{\rho }\,Q_{\sigma }\,
      \eta_{\alpha\beta }}\over {2\,{(k_2\cdot Q)^2}}} \cr & -
  {{{k_1}_{\nu }\,Q_{\beta }\,Q_{\rho }\,Q_{\sigma }\,
      \eta_{\alpha\mu }}\over {{k_1}\cdot Q}} -
  {{{k_2}_{\nu }\,Q_{\beta }\,Q_{\rho }\,Q_{\sigma }\,
      \eta_{\alpha\mu }}\over {{k_1}\cdot Q}} +
  {{{k_2}_{\nu }\,Q_{\beta }\,Q_{\rho }\,Q_{\sigma }\,
      \eta_{\alpha\mu }}\over {{k_2}\cdot Q}} \cr & +
  {{{k_1}_{\beta }\,Q_{\nu }\,Q_{\rho }\,Q_{\sigma }\,
      \eta_{\alpha\mu }}\over {{k_1}\cdot Q}} -
  {{{k_1}_{\beta }\,Q_{\nu }\,Q_{\rho }\,Q_{\sigma }\,
      \eta_{\alpha\mu }}\over {{k_2}\cdot Q}} -
  {{{k_2}_{\beta }\,Q_{\nu }\,Q_{\rho }\,Q_{\sigma }\,
      \eta_{\alpha\mu }}\over {{k_2}\cdot Q}} \cr & +
  {{4\,{k_1}\cdot {k_2}\,Q_{\beta }\,Q_{\nu }\,Q_{\rho }\,Q_{\sigma }\,
      \eta_{\alpha\mu }}\over {{k_1}\cdot Q\,{k_2}\cdot Q}} -
  {{{k_1}_{\mu }\,Q_{\beta }\,Q_{\rho }\,Q_{\sigma }\,
      \eta_{\alpha\nu }}\over {{k_1}\cdot Q}} -
  {{{k_2}_{\mu }\,Q_{\beta }\,Q_{\rho }\,Q_{\sigma }\,
      \eta_{\alpha\nu }}\over {{k_1}\cdot Q}} \cr & +
  {{{k_2}_{\mu }\,Q_{\beta }\,Q_{\rho }\,Q_{\sigma }\,
      \eta_{\alpha\nu }}\over {{k_2}\cdot Q}} +
  {{{k_1}_{\beta }\,Q_{\mu }\,Q_{\rho }\,Q_{\sigma }\,
      \eta_{\alpha\nu }}\over {{k_1}\cdot Q}} -
  {{{k_1}_{\beta }\,Q_{\mu }\,Q_{\rho }\,Q_{\sigma }\,
      \eta_{\alpha\nu }}\over {{k_2}\cdot Q}} \cr & -
  {{{k_2}_{\beta }\,Q_{\mu }\,Q_{\rho }\,Q_{\sigma }\,
      \eta_{\alpha\nu }}\over {{k_2}\cdot Q}} +
  {{4\,{k_1}\cdot {k_2}\,Q_{\beta }\,Q_{\mu }\,Q_{\rho }\,Q_{\sigma }\,
      \eta_{\alpha\nu }}\over {{k_1}\cdot Q\,{k_2}\cdot Q}} -
  {{{k_2}_{\sigma }\,Q_{\beta }\,Q_{\mu }\,Q_{\nu }\,
      \eta_{\alpha\rho }}\over {{k_1}\cdot Q}} \cr & -
  {{{k_2}_{\sigma }\,Q_{\beta }\,Q_{\mu }\,Q_{\nu }\,
      \eta_{\alpha\rho }}\over {{k_2}\cdot Q}} -
  {{{k_1}_{\beta }\,Q_{\mu }\,Q_{\nu }\,Q_{\sigma }\,
      \eta_{\alpha\rho }}\over {{k_1}\cdot Q}} +
  {{{k_2}_{\beta }\,Q_{\mu }\,Q_{\nu }\,Q_{\sigma }\,
      \eta_{\alpha\rho }}\over {{k_2}\cdot Q}} \cr & -
  {{4\,k_1^2\,Q_{\beta }\,Q_{\mu }\,Q_{\nu }\,Q_{\sigma }\,
      \eta_{\alpha\rho }}\over {{k_1}\cdot Q\,{k_2}\cdot Q}} -
  {{4\,{k_1}\cdot {k_2}\,Q_{\beta }\,Q_{\mu }\,Q_{\nu }\,Q_{\sigma }\,
      \eta_{\alpha\rho }}\over {{k_1}\cdot Q\,{k_2}\cdot Q}} -
  {{{k_2}_{\rho }\,Q_{\beta }\,Q_{\mu }\,Q_{\nu }\,
      \eta_{\alpha\sigma }}\over {{k_1}\cdot Q}} \cr }$$
\vfill\eject
$$\eqalign{& -
  {{{k_2}_{\rho }\,Q_{\beta }\,Q_{\mu }\,Q_{\nu }\,
      \eta_{\alpha\sigma }}\over {{k_2}\cdot Q}} -
  {{{k_1}_{\beta }\,Q_{\mu }\,Q_{\nu }\,Q_{\rho }\,
      \eta_{\alpha\sigma }}\over {{k_1}\cdot Q}} +
  {{{k_2}_{\beta }\,Q_{\mu }\,Q_{\nu }\,Q_{\rho }\,
      \eta_{\alpha\sigma }}\over {{k_2}\cdot Q}} \cr & -
  {{4\,k_1^2\,Q_{\beta }\,Q_{\mu }\,Q_{\nu }\,Q_{\rho }\,
      \eta_{\alpha\sigma }}\over {{k_1}\cdot Q\,{k_2}\cdot Q}} -
  {{4\,{k_1}\cdot {k_2}\,Q_{\beta }\,Q_{\mu }\,Q_{\nu }\,Q_{\rho }\,
      \eta_{\alpha\sigma }}\over {{k_1}\cdot Q\,{k_2}\cdot Q}} +
  2\,Q_{\rho }\,Q_{\sigma }\,\eta_{\alpha\mu }\,\eta_{\beta\nu } \cr & +
  {{{k_1}\cdot Q\,Q_{\mu }\,Q_{\sigma }\,\eta_{\alpha\rho }\,
      \eta_{\beta\nu }}\over {{k_2}\cdot Q}} +
  {{{k_1}\cdot Q\,Q_{\nu }\,Q_{\sigma }\,\eta_{\alpha\mu }\,
      \eta_{\beta\rho }}\over {{k_2}\cdot Q}} +
  {{{k_1}\cdot Q\,Q_{\nu }\,Q_{\rho }\,\eta_{\alpha\mu }\,
      \eta_{\beta\sigma }}\over {{k_2}\cdot Q}} \cr & +
  {{{k_1}\cdot Q\,Q_{\mu }\,Q_{\rho }\,\eta_{\alpha\nu }\,
      \eta_{\beta\sigma }}\over {{k_2}\cdot Q}} +
  6\,Q_{\mu }\,Q_{\nu }\,\eta_{\alpha\rho }\,\eta_{\beta\sigma } -
  {{4\,{k_1}\cdot Q\,Q_{\mu }\,Q_{\nu }\,\eta_{\alpha\rho }\,
      \eta_{\beta\sigma }}\over {{k_2}\cdot Q}} \cr & -
  {{5\,{k_1}_{\rho }\,Q_{\alpha }\,Q_{\beta }\,Q_{\sigma }\,
      \eta_{\mu\nu }}\over {{k_1}\cdot Q}} +
  {{2\,{k_1}_{\rho }\,Q_{\alpha }\,Q_{\beta }\,Q_{\sigma }\,
      \eta_{\mu\nu }}\over {{k_2}\cdot Q}} +
  {{2\,{k_2}_{\rho }\,Q_{\alpha }\,Q_{\beta }\,Q_{\sigma }\,
      \eta_{\mu\nu }}\over {{k_1}\cdot Q}} \cr & -
  {{5\,{k_2}_{\rho }\,Q_{\alpha }\,Q_{\beta }\,Q_{\sigma }\,
      \eta_{\mu\nu }}\over {{k_2}\cdot Q}} -
  {{5\,{k_1}_{\alpha }\,Q_{\beta }\,Q_{\rho }\,Q_{\sigma }\,
      \eta_{\mu\nu }}\over {{k_1}\cdot Q}} +
  {{2\,{k_1}_{\alpha }\,Q_{\beta }\,Q_{\rho }\,Q_{\sigma }\,
      \eta_{\mu\nu }}\over {{k_2}\cdot Q}} \cr & +
  {{3\,{k_2}_{\alpha }\,Q_{\beta }\,Q_{\rho }\,Q_{\sigma }\,
      \eta_{\mu\nu }}\over {{k_2}\cdot Q}} +
  {{5\,k_1^2\,Q_{\alpha }\,Q_{\beta }\,Q_{\rho }\,Q_{\sigma }\,
      \eta_{\mu\nu }}\over {2\,{(k_1\cdot Q)^2}}} +
  {{k_2^2\,Q_{\alpha }\,Q_{\beta }\,Q_{\rho }\,Q_{\sigma }\,
      \eta_{\mu\nu }}\over {{(k_2\cdot Q)^2}}} \cr & +
  Q_{\rho }\,Q_{\sigma }\,\eta_{\alpha\beta }\,\eta_{\mu\nu } -
  {{2\,{k_1}\cdot Q\,Q_{\beta }\,Q_{\sigma }\,\eta_{\alpha\rho }\,
      \eta_{\mu\nu }}\over {{k_2}\cdot Q}} -
  {{{k_2}\cdot Q\,Q_{\beta }\,Q_{\sigma }\,\eta_{\alpha\rho }\,
      \eta_{\mu\nu }}\over {{k_1}\cdot Q}} \cr & -
  {{2\,{k_1}\cdot Q\,Q_{\beta }\,Q_{\rho }\,\eta_{\alpha\sigma }\,
      \eta_{\mu\nu }}\over {{k_2}\cdot Q}} -
  {{{k_2}\cdot Q\,Q_{\beta }\,Q_{\rho }\,\eta_{\alpha\sigma }\,
      \eta_{\mu\nu }}\over {{k_1}\cdot Q}} -
  {{{k_1}_{\sigma }\,Q_{\alpha }\,Q_{\beta }\,Q_{\nu }\,
      \eta_{\mu\rho }}\over {{k_1}\cdot Q}} \cr & -
  {{{k_1}_{\sigma }\,Q_{\alpha }\,Q_{\beta }\,Q_{\nu }\,
      \eta_{\mu\rho }}\over {{k_2}\cdot Q}} +
  {{{k_1}_{\nu }\,Q_{\alpha }\,Q_{\beta }\,Q_{\sigma }\,
      \eta_{\mu\rho }}\over {{k_1}\cdot Q}} -
  {{{k_2}_{\nu }\,Q_{\alpha }\,Q_{\beta }\,Q_{\sigma }\,
      \eta_{\mu\rho }}\over {{k_2}\cdot Q}} \cr & -
  {{4\,{k_1}\cdot {k_2}\,Q_{\alpha }\,Q_{\beta }\,Q_{\nu }\,
      Q_{\sigma }\,\eta_{\mu\rho }}\over {{k_1}\cdot Q\,{k_2}\cdot Q}} -
  {{4\,k_2^2\,Q_{\alpha }\,Q_{\beta }\,Q_{\nu }\,Q_{\sigma }\,
      \eta_{\mu\rho }}\over {{k_1}\cdot Q\,{k_2}\cdot Q}} -
  {{{k_1}\cdot Q\,Q_{\nu }\,Q_{\sigma }\,\eta_{\alpha\beta }\,
      \eta_{\mu\rho }}\over {{k_2}\cdot Q}} \cr & -
  {{2\,{k_2}\cdot Q\,Q_{\nu }\,Q_{\sigma }\,\eta_{\alpha\beta }\,
      \eta_{\mu\rho }}\over {{k_1}\cdot Q}} +
  {{2\,{k_2}\cdot Q\,Q_{\beta }\,Q_{\sigma }\,\eta_{\alpha\nu }\,
      \eta_{\mu\rho }}\over {{k_1}\cdot Q}} +
  {{2\,{k_1}\cdot Q\,Q_{\beta }\,Q_{\nu }\,\eta_{\alpha\sigma }\,
      \eta_{\mu\rho }}\over {{k_2}\cdot Q}} \cr & +
  {{2\,{k_2}\cdot Q\,Q_{\beta }\,Q_{\nu }\,\eta_{\alpha\sigma }\,
      \eta_{\mu\rho }}\over {{k_1}\cdot Q}} -
  {{{k_1}_{\rho }\,Q_{\alpha }\,Q_{\beta }\,Q_{\nu }\,
      \eta_{\mu\sigma }}\over {{k_1}\cdot Q}} -
  {{{k_1}_{\rho }\,Q_{\alpha }\,Q_{\beta }\,Q_{\nu }\,
      \eta_{\mu\sigma }}\over {{k_2}\cdot Q}} \cr & +
  {{{k_1}_{\nu }\,Q_{\alpha }\,Q_{\beta }\,Q_{\rho }\,
      \eta_{\mu\sigma }}\over {{k_1}\cdot Q}} -
  {{{k_2}_{\nu }\,Q_{\alpha }\,Q_{\beta }\,Q_{\rho }\,
      \eta_{\mu\sigma }}\over {{k_2}\cdot Q}} -
  {{4\,{k_1}\cdot {k_2}\,Q_{\alpha }\,Q_{\beta }\,Q_{\nu }\,Q_{\rho }\,
      \eta_{\mu\sigma }}\over {{k_1}\cdot Q\,{k_2}\cdot Q}} \cr & -
  {{4\,k_2^2\,Q_{\alpha }\,Q_{\beta }\,Q_{\nu }\,Q_{\rho }\,
      \eta_{\mu\sigma }}\over {{k_1}\cdot Q\,{k_2}\cdot Q}} -
  {{{k_1}\cdot Q\,Q_{\nu }\,Q_{\rho }\,\eta_{\alpha\beta }\,
      \eta_{\mu\sigma }}\over {{k_2}\cdot Q}} -
  {{2\,{k_2}\cdot Q\,Q_{\nu }\,Q_{\rho }\,\eta_{\alpha\beta }\,
      \eta_{\mu\sigma }}\over {{k_1}\cdot Q}} \cr & +
  {{2\,{k_1}\cdot Q\,Q_{\beta }\,Q_{\nu }\,\eta_{\alpha\rho }\,
      \eta_{\mu\sigma }}\over {{k_2}\cdot Q}} +
  {{2\,{k_2}\cdot Q\,Q_{\beta }\,Q_{\nu }\,\eta_{\alpha\rho }\,
      \eta_{\mu\sigma }}\over {{k_1}\cdot Q}} +
  {{2\,{k_2}\cdot Q\,Q_{\beta }\,Q_{\sigma }\,\eta_{\alpha\mu }\,
      \eta_{\nu\rho }}\over {{k_1}\cdot Q}} \cr & +
  6\,Q_{\alpha }\,Q_{\beta }\,\eta_{\mu\rho }\,\eta_{\nu\sigma } -
  {{4\,{k_2}\cdot Q\,Q_{\alpha }\,Q_{\beta }\,\eta_{\mu\rho }\,
      \eta_{\nu\sigma }}\over {{k_1}\cdot Q}} -
  {{3\,{k_1}_{\mu }\,Q_{\alpha }\,Q_{\beta }\,Q_{\nu }\,
      \eta_{\rho\sigma }}\over {{k_1}\cdot Q}} \cr & +
  {{2\,{k_2}_{\mu }\,Q_{\alpha }\,Q_{\beta }\,Q_{\nu }\,
      \eta_{\rho\sigma }}\over {{k_1}\cdot Q}} -
  {{5\,{k_2}_{\mu }\,Q_{\alpha }\,Q_{\beta }\,Q_{\nu }\,
      \eta_{\rho\sigma }}\over {{k_2}\cdot Q}} -
  {{5\,{k_1}_{\alpha }\,Q_{\beta }\,Q_{\mu }\,Q_{\nu }\,
      \eta_{\rho\sigma }}\over {{k_1}\cdot Q}} \cr & +
  {{2\,{k_1}_{\alpha }\,Q_{\beta }\,Q_{\mu }\,Q_{\nu }\,
      \eta_{\rho\sigma }}\over {{k_2}\cdot Q}} -
  {{3\,{k_2}_{\alpha }\,Q_{\beta }\,Q_{\mu }\,Q_{\nu }\,
      \eta_{\rho\sigma }}\over {{k_2}\cdot Q}} +
  {{5\,k_1^2\,Q_{\alpha }\,Q_{\beta }\,Q_{\mu }\,Q_{\nu }\,
      \eta_{\rho\sigma }}\over {2\,{(k_1\cdot Q)^2}}} \cr }$$
\vfill\eject
$$\eqalign{& +
  {{5\,k_2^2\,Q_{\alpha }\,Q_{\beta }\,Q_{\mu }\,Q_{\nu }\,
      \eta_{\rho\sigma }}\over {2\,{(k_2\cdot Q)^2}}} +
  2\,Q_{\mu }\,Q_{\nu }\,\eta_{\alpha\beta }\,\eta_{\rho\sigma } -
  {{{k_1}\cdot Q\,Q_{\beta }\,Q_{\nu }\,\eta_{\alpha\mu }\,
      \eta_{\rho\sigma }}\over {{k_2}\cdot Q}} \cr & -
  {{{k_2}\cdot Q\,Q_{\beta }\,Q_{\nu }\,\eta_{\alpha\mu }\,
      \eta_{\rho\sigma }}\over {{k_1}\cdot Q}} -
  {{{k_1}\cdot Q\,Q_{\beta }\,Q_{\mu }\,\eta_{\alpha\nu }\,
      \eta_{\rho\sigma }}\over {{k_2}\cdot Q}} -
  {{{k_2}\cdot Q\,Q_{\beta }\,Q_{\mu }\,\eta_{\alpha\nu }\,
      \eta_{\rho\sigma }}\over {{k_1}\cdot Q}} \cr & +
  2\,Q_{\alpha }\,Q_{\beta }\,\eta_{\mu\nu }\,\eta_{\rho\sigma }\cr & +
(symmetrization~under~(\alpha\leftrightarrow\beta),~
                        (\mu\leftrightarrow\nu),~
                        (\rho\leftrightarrow\sigma))\left.\right] \cr & +
\left[(k_1,~\alpha,~\beta)\leftrightarrow(k_3,~\rho,~\sigma)\right]+
\left[(k_2,~\mu,~\nu)\leftrightarrow(k_3,~\rho,~\sigma)\right],}\eapb$$
$$\eqalign{&
{1\over \kappa^3}\hat\Gamma^{3b}_{(\alpha\beta)(\mu\nu)(\rho\sigma)}
(k_1,k_2,k_3,Q)= \cr &
\left[\right.
-{{{k_1}_{\rho }\,{k_1}_{\sigma }\,Q_{\alpha }\,Q_{\beta }\,Q_{\mu }\,
       Q_{\nu }}\over {{k_1}\cdot Q\,{k_2}\cdot Q}} -
  {{10\,{k_1}_{\rho }\,{k_2}_{\sigma }\,Q_{\alpha }\,Q_{\beta }\,
      Q_{\mu }\,Q_{\nu }}\over {{k_1}\cdot Q\,{k_2}\cdot Q}} -
  {{{k_2}_{\rho }\,{k_2}_{\sigma }\,Q_{\alpha }\,Q_{\beta }\,Q_{\mu }\,
      Q_{\nu }}\over {{k_1}\cdot Q\,{k_2}\cdot Q}} \cr & -
  {{10\,{k_1}_{\rho }\,{k_2}_{\mu }\,Q_{\alpha }\,Q_{\beta }\,Q_{\nu }\,
      Q_{\sigma }}\over {{k_1}\cdot Q\,{k_2}\cdot Q}} +
  {{6\,{k_2}_{\mu }\,{k_2}_{\rho }\,Q_{\alpha }\,Q_{\beta }\,Q_{\nu }\,
      Q_{\sigma }}\over {{k_1}\cdot Q\,{k_2}\cdot Q}} +
  {{6\,{k_1}_{\alpha }\,{k_1}_{\rho }\,Q_{\beta }\,Q_{\mu }\,Q_{\nu }\,
      Q_{\sigma }}\over {{k_1}\cdot Q\,{k_2}\cdot Q}} \cr & -
  {{10\,{k_1}_{\alpha }\,{k_2}_{\rho }\,Q_{\beta }\,Q_{\mu }\,Q_{\nu }\,
      Q_{\sigma }}\over {{k_1}\cdot Q\,{k_2}\cdot Q}} +
  {{4\,{k_1}_{\rho }\,k_2^2\,Q_{\alpha }\,Q_{\beta }\,Q_{\mu }\,
      Q_{\nu }\,Q_{\sigma }}\over {{k_1}\cdot Q\,{(k_2\cdot Q)^2}}}
   - {{4\,k_1^2\,{k_1}_{\rho }\,Q_{\alpha }\,Q_{\beta }\,Q_{\mu }\,
      Q_{\nu }\,Q_{\sigma }}\over {{(k_1\cdot Q)^2}\,{k_2}\cdot Q}}
   \cr & -
{{4\,k_2^2\,{k_2}_{\rho }\,Q_{\alpha }\,Q_{\beta }\,Q_{\mu }\,
      Q_{\nu }\,Q_{\sigma }}\over {{k_1}\cdot Q\,{(k_2\cdot Q)^2}}}
   + {{4\,k_1^2\,{k_2}_{\rho }\,Q_{\alpha }\,Q_{\beta }\,Q_{\mu }\,
      Q_{\nu }\,Q_{\sigma }}\over {{(k_1\cdot Q)^2}\,{k_2}\cdot Q}}
   - {{{k_2}_{\mu }\,{k_2}_{\nu }\,Q_{\alpha }\,Q_{\beta }\,Q_{\rho }\,
      Q_{\sigma }}\over {{k_1}\cdot Q\,{k_2}\cdot Q}} \cr & -
  {{6\,{k_1}_{\alpha }\,{k_2}_{\mu }\,Q_{\beta }\,Q_{\nu }\,Q_{\rho }\,
      Q_{\sigma }}\over {{k_1}\cdot Q\,{k_2}\cdot Q}} -
  {{4\,k_2^2\,{k_2}_{\mu }\,Q_{\alpha }\,Q_{\beta }\,Q_{\nu }\,
      Q_{\rho }\,Q_{\sigma }}\over {{k_1}\cdot Q\,{(k_2\cdot Q)^2}}}
    + {{4\,k_1^2\,{k_2}_{\mu }\,Q_{\alpha }\,Q_{\beta }\,Q_{\nu }\,
      Q_{\rho }\,Q_{\sigma }}\over {{(k_1\cdot Q)^2}\,{k_2}\cdot Q}}
    \cr & -
 {{{k_1}_{\alpha }\,{k_1}_{\beta }\,Q_{\mu }\,Q_{\nu }\,Q_{\rho }\,
      Q_{\sigma }}\over {{k_1}\cdot Q\,{k_2}\cdot Q}} +
  {{4\,{k_1}_{\alpha }\,k_2^2\,Q_{\beta }\,Q_{\mu }\,Q_{\nu }\,
      Q_{\rho }\,Q_{\sigma }}\over {{k_1}\cdot Q\,{(k_2\cdot Q)^2}}}
    - {{4\,k_1^2\,{k_1}_{\alpha }\,Q_{\beta }\,Q_{\mu }\,Q_{\nu }\,
      Q_{\rho }\,Q_{\sigma }}\over {{{{k_1}\cdot Q}^2}\,{k_2}\cdot Q}}
    \cr & +
{{2\,{k_2^4}\,Q_{\alpha }\,Q_{\beta }\,Q_{\mu }\,Q_{\nu }\,
      Q_{\rho }\,Q_{\sigma }}\over {{k_1}\cdot Q\,{(k_2\cdot Q)^3}}}
    - {{2\,k_1^2\,k_2^2\,Q_{\alpha }\,Q_{\beta }\,Q_{\mu }\,
      Q_{\nu }\,Q_{\rho }\,Q_{\sigma }}\over
    {{(k_1\cdot Q)^2}\,{(k_2\cdot Q)^2}}} +
  {{2\,{k_1^4}\,Q_{\alpha }\,Q_{\beta }\,Q_{\mu }\,Q_{\nu }\,
      Q_{\rho }\,Q_{\sigma }}\over {{(k_1\cdot Q)^3}\,{k_2}\cdot Q}}\cr & +
(symmetrization~under~(\alpha\leftrightarrow\beta),~
                        (\mu\leftrightarrow\nu),~
                        (\rho\leftrightarrow\sigma))\left.\right]\cr & +
\left[(k_1,~\alpha,~\beta)\leftrightarrow(k_3,~\rho,~\sigma)\right]+
\left[(k_2,~\mu,~\nu)\leftrightarrow(k_3,~\rho,~\sigma)\right],}\eapb$$

\vfill\eject
$$\eqalign{&{1\over \kappa^3}\hat\Gamma^{3c}_{(\alpha\beta)(\mu\nu)
(\rho\sigma)}
(k_1,k_2,k_3,Q)= \cr & +
\left[\right.
{{4\,{k_1}_{\rho }\,Q_{\mu }\,Q_{\nu }\,Q_{\sigma }\,
      \eta_{\alpha\beta }}\over {{k_1}\cdot Q}} +
  {{4\,{k_1}_{\mu }\,Q_{\nu }\,Q_{\rho }\,Q_{\sigma }\,
      \eta_{\alpha\beta }}\over {{k_1}\cdot Q}} -
  {{2\,k_1^2\,Q_{\mu }\,Q_{\nu }\,Q_{\rho }\,Q_{\sigma }\,
      \eta_{\alpha\beta }}\over {{(k_1\cdot Q)^2}}} \cr & -
  8\,Q_{\rho }\,Q_{\sigma }\,\eta_{\alpha\mu }\,\eta_{\beta\nu } -
  8\,Q_{\mu }\,Q_{\nu }\,\eta_{\alpha\rho }\,\eta_{\beta\sigma } +
  {{10\,{k_1}_{\rho }\,Q_{\alpha }\,Q_{\beta }\,Q_{\sigma }\,
      \eta_{\mu\nu }}\over {{k_1}\cdot Q}} \cr & +
  {{10\,{k_1}_{\alpha }\,Q_{\beta }\,Q_{\rho }\,Q_{\sigma }\,
      \eta_{\mu\nu }}\over {{k_1}\cdot Q}} -
  {{5\,k_1^2\,Q_{\alpha }\,Q_{\beta }\,Q_{\rho }\,Q_{\sigma }\,
      \eta_{\mu\nu }}\over {{(k_1\cdot Q)^2}}} -
  3\,Q_{\rho }\,Q_{\sigma }\,\eta_{\alpha\beta }\,\eta_{\mu\nu } \cr & -
  Q_{\beta }\,Q_{\sigma }\,\eta_{\alpha\rho }\,\eta_{\mu\nu } -
  Q_{\beta }\,Q_{\rho }\,\eta_{\alpha\sigma }\,\eta_{\mu\nu } -
  2\,Q_{\nu }\,Q_{\sigma }\,\eta_{\alpha\beta }\,\eta_{\mu\rho } \cr & +
  2\,Q_{\beta }\,Q_{\sigma }\,\eta_{\alpha\nu }\,\eta_{\mu\rho } +
  2\,Q_{\beta }\,Q_{\nu }\,\eta_{\alpha\sigma }\,\eta_{\mu\rho } -
  2\,Q_{\nu }\,Q_{\rho }\,\eta_{\alpha\beta }\,\eta_{\mu\sigma } \cr & +
  2\,Q_{\beta }\,Q_{\nu }\,\eta_{\alpha\rho }\,\eta_{\mu\sigma } +
  2\,Q_{\beta }\,Q_{\sigma }\,\eta_{\alpha\mu }\,\eta_{\nu\rho } -
  20\,Q_{\alpha }\,Q_{\beta }\,\eta_{\mu\rho }\,\eta_{\nu\sigma } \cr & +
  {{10\,{k_1}_{\mu }\,Q_{\alpha }\,Q_{\beta }\,Q_{\nu }\,
      \eta_{\rho\sigma }}\over {{k_1}\cdot Q}} +
  {{10\,{k_1}_{\alpha }\,Q_{\beta }\,Q_{\mu }\,Q_{\nu }\,
      \eta_{\rho\sigma }}\over {{k_1}\cdot Q}} -
  {{5\,k_1^2\,Q_{\alpha }\,Q_{\beta }\,Q_{\mu }\,Q_{\nu }\,
      \eta_{\rho\sigma }}\over {{(k_1\cdot Q)^2}}} \cr & -
  3\,Q_{\mu }\,Q_{\nu }\,\eta_{\alpha\beta }\,\eta_{\rho\sigma } -
  Q_{\beta }\,Q_{\nu }\,\eta_{\alpha\mu }\,\eta_{\rho\sigma } -
  Q_{\beta }\,Q_{\mu }\,\eta_{\alpha\nu }\,\eta_{\rho\sigma } \cr & -
  2\,Q_{\alpha }\,Q_{\beta }\,\eta_{\mu\nu }\,\eta_{\rho\sigma }\cr & +
(symmetrization~under~(\alpha\leftrightarrow\beta),~
                        (\mu\leftrightarrow\nu),~
                        (\rho\leftrightarrow\sigma))\left.\right]\cr & +
\left[ciclic~permutations~of~(k_1,~\alpha,~\beta)~(k_2,~\mu,~\nu)~
and~(k_3,~\rho,~\sigma)\right],\cr}\eapb$$
$$\eqalign{&{1\over \kappa^3}
\hat\Gamma^{3d}_{(\alpha\beta)(\mu\nu)(\rho\sigma)}
(k_1,k_2,k_3,Q)= \cr &
+{{9\,Q_{\rho }\,Q_{\sigma }\,\eta_{\alpha\nu }\,\eta_{\beta\mu }}} +
{{9\,Q_{\rho }\,Q_{\sigma }\,\eta_{\alpha\mu }\,\eta_{\beta\nu }}} +
{{9\,Q_{\mu }\,Q_{\nu }\,\eta_{\alpha\sigma }\,\eta_{\beta\rho }}} \cr & +
{{9\,Q_{\mu }\,Q_{\nu }\,\eta_{\alpha\rho }\,\eta_{\beta\sigma }}} +
{{3\,Q_{\rho }\,Q_{\sigma }\,\eta_{\alpha  \beta }\,\eta_{\mu\nu }}} +
{{9\,Q_{\alpha }\,Q_{\beta }\,\eta_{\mu\sigma }\,\eta_{\nu\rho }}} \cr & +
{{9\,Q_{\alpha }\,Q_{\beta }\,\eta_{\mu\rho }\,\eta_{\nu\sigma }}} +
{{3\,Q_{\mu }\,Q_{\nu }\,\eta_{\alpha  \beta }\,\eta_{\rho\sigma }}} +
{{3\,Q_{\alpha }\,Q_{\beta }\,\eta_{\mu\nu }\,
\eta_{\rho\sigma }}},\cr}\eapb$$
The sum of these contributions gives, with the help of the eikonal identity
(3.2), the following result:
\vfill\eject
$$\eqalign{&{1\over \kappa^3}
\hat\Gamma^{3}_{(\alpha\beta)(\mu\nu)(\rho\sigma)}(k_1,k_2,k_3,Q)= \cr &
2\left[\right.
{{{k_1}_{\rho }\,{k_2}_{\sigma }\,Q_{\alpha }\,Q_{\beta }\,Q_{\mu }\,
      Q_{\nu }}\over {{k_1}\cdot Q\,{k_2}\cdot Q}} +
  {{{k_1}_{\rho }\,{k_2}_{\mu }\,Q_{\alpha }\,Q_{\beta }\,Q_{\nu }\,
      Q_{\sigma }}\over {{k_1}\cdot Q\,{k_2}\cdot Q}} -
  {{{k_2}_{\mu }\,{k_2}_{\rho }\,Q_{\alpha }\,Q_{\beta }\,Q_{\nu }\,
      Q_{\sigma }}\over {{k_1}\cdot Q\,{k_2}\cdot Q}} \cr & -
  {{{k_1}_{\alpha }\,{k_1}_{\rho }\,Q_{\beta }\,Q_{\mu }\,Q_{\nu }\,
      Q_{\sigma }}\over {{k_1}\cdot Q\,{k_2}\cdot Q}} +
  {{{k_1}_{\alpha }\,{k_2}_{\rho }\,Q_{\beta }\,Q_{\mu }\,Q_{\nu }\,
      Q_{\sigma }}\over {{k_1}\cdot Q\,{k_2}\cdot Q}} -
  {{{k_1}_{\rho }\,k_2^2\,Q_{\alpha }\,Q_{\beta }\,Q_{\mu }\,
      Q_{\nu }\,Q_{\sigma }}\over
    {2\,{k_1}\cdot Q\,{(k_2\cdot Q)^2}}} \cr & +
  {{k_1^2\,{k_1}_{\rho }\,Q_{\alpha }\,Q_{\beta }\,Q_{\mu }\,
      Q_{\nu }\,Q_{\sigma }}\over
    {2\,{(k_1\cdot Q)^2}\,{k_2}\cdot Q}} +
  {{k_2^2\,{k_2}_{\rho }\,Q_{\alpha }\,Q_{\beta }\,Q_{\mu }\,
      Q_{\nu }\,Q_{\sigma }}\over
    {2\,{k_1}\cdot Q\,{(k_2\cdot Q)^2}}} -
  {{k_1^2\,{k_2}_{\rho }\,Q_{\alpha }\,Q_{\beta }\,Q_{\mu }\,
      Q_{\nu }\,Q_{\sigma }}\over
    {2\,{(k_1\cdot Q)^2}\,{k_2}\cdot Q}} \cr & +
  {{{k_1}_{\alpha }\,{k_2}_{\mu }\,Q_{\beta }\,Q_{\nu }\,Q_{\rho }\,
      Q_{\sigma }}\over {{k_1}\cdot Q\,{k_2}\cdot Q}} +
  {{k_2^2\,{k_2}_{\mu }\,Q_{\alpha }\,Q_{\beta }\,Q_{\nu }\,
      Q_{\rho }\,Q_{\sigma }}\over
    {2\,{k_1}\cdot Q\,{(k_2\cdot Q)^2}}} -
  {{k_1^2\,{k_2}_{\mu }\,Q_{\alpha }\,Q_{\beta }\,Q_{\nu }\,
      Q_{\rho }\,Q_{\sigma }}\over
    {2\,{(k_1\cdot Q)^2}\,{k_2}\cdot Q}} \cr & -
  {{{k_1}_{\alpha }\,k_2^2\,Q_{\beta }\,Q_{\mu }\,Q_{\nu }\,
      Q_{\rho }\,Q_{\sigma }}\over
    {2\,{k_1}\cdot Q\,{(k_2\cdot Q)^2}}} +
  {{k_1^2\,{k_1}_{\alpha }\,Q_{\beta }\,Q_{\mu }\,Q_{\nu }\,
      Q_{\rho }\,Q_{\sigma }}\over
    {2\,{(k_1\cdot Q)^2}\,{k_2}\cdot Q}} -
  {{{k_2^4}\,Q_{\alpha }\,Q_{\beta }\,Q_{\mu }\,Q_{\nu }\,
      Q_{\rho }\,Q_{\sigma }}\over
    {4\,{k_1}\cdot Q\,{(k_2\cdot Q)^3}}} \cr & +
  {{k_1^2\,k_2^2\,Q_{\alpha }\,Q_{\beta }\,Q_{\mu }\,Q_{\nu }\,
      Q_{\rho }\,Q_{\sigma }}\over
    {4\,{(k_1\cdot Q)^2}\,{(k_2\cdot Q)^2}}} -
  {{{k_1^4}\,Q_{\alpha }\,Q_{\beta }\,Q_{\mu }\,Q_{\nu }\,
      Q_{\rho }\,Q_{\sigma }}\over
    {4\,{(k_1\cdot Q)^3}\,{k_2}\cdot Q}}\cr & +
(symmetrization~under~(\alpha\leftrightarrow\beta),~
                        (\mu\leftrightarrow\nu),~
                        (\rho\leftrightarrow\sigma))\left.\right]\cr & +
2 \left[(k_1~,\alpha,~\beta)\leftrightarrow(k_3,~\rho,~\sigma)\right]+
2 \left[(k_2,~\mu,~\nu)\leftrightarrow(k_3,~\rho,~\sigma)\right].\cr}\eapb$$
Using this result, together with the eikonal identity, we arrive after a
straightforward calculation at the expression given by equation (3.5).

\vfill\eject

\beginsection \centerline{REFERENCES}

\item{$[1]$}D. I. Gross, M. J. Perry, and L. G. Yaffe, Phys. Rev. D {\bf 25},
330 (1982); Y. Kikuchi, T. Moriya, and T. Tsukahara, Phys. Rev. D {\bf 29},
2220 (1984); P. S. Gribosky, J. F. Donoghue, and B. R. Holstein, Ann. Phys.
(N.Y.) {\bf 190}, 149 (1989).
\item{$[2]$}A. Rebhan, Nucl. Phys. {\bf B351}, 706 (1991).
\item{$[3]$}J. Frenkel and J. C. Taylor, Z. Phys. C {\bf 49}, 515 (1991);
F. T. Brandt, J. Frenkel, and J. C. Taylor, Nucl. Phys. {\bf B374}, 169 (1992).
\item{$[4]$}E. Braaten and R. D. Pisarski, Nucl. Phys. {\bf B337}, 569 (1990);
{\it ibid.} {\bf B339}, 310 (1990).
\item{$[5]$}J. Frenkel and J. C. Taylor, Nucl. Phys. {\bf B374}, 156 (1992).
\item{$[6]$}D. M. Capper, G. Leibbrandt, and M. Ram\'on Medrano, Phys. Rev.
D {\bf 8}, 4320 (1973).
\item{$[7]$}J. I. Kapusta, {\it Finite Temperature Field Theory} (Cambridge
University Press, Cambridge, England 1989).

\vfill\eject

\beginsection \centerline{FIGURE CAPTIONS}

\pa
\item{{\it Fig.1}~--} Lowest order contributions to the thermal graviton
self-energy. Wavy lines denote gravitons and dashed lines represent ghost
particles.
\item{{\it Fig.2}~--} The forward scattering diagrams corresponding to Fig. 1.
Crossed graphs with {$k\rightarrow -k$} are to be understood.
\item{{\it Fig.3}~--} Feynman diagrams contributing to the thermal 3-graviton
vertex function. Graphs obtained by permutations of external gravitons in (c)
are to be understood.
\item{{\it Fig.4}~--}  Examples of forward scattering graphs connected with
Fig. 3. Diagrams obtained by permutations of external gravitons should be
understood.

\vfill\eject
\end